\documentclass{article}
\usepackage{spconf,amsmath,graphicx}

\usepackage{multirow}
\usepackage{cite}
\usepackage[colorlinks]{hyperref}
\usepackage[table,xcdraw]{xcolor}
\usepackage{subfig}
\usepackage[capitalise]{cleveref}
\usepackage[printonlyused]{acronym} % include acronym file (exclude from the list)
\acrodef {BCE}[BCE]{binary cross entropy}
\acrodef {MAE}[MAE]{mean average error}
\acrodef {RMSE}[RMSE]{Root mean square error}
\acrodef {MSE}[MSE]{mean squared error}
\acrodef {NMS}[NMS]{non-maximum suppression}
\acrodef {DNN}[DNNs]{deep neural networks}
\acrodef {TDOA}[TDOA]{time difference of arrival}
\acrodef {DOA}[DOA]{direction of arrival}
\acrodef {SSL}[SSL]{Sound source localization}
\acrodef {STFT}[STFT]{short-time Fourier transform}
\acrodef {tgrid}[TG-rep]{Tight grid}
\acrodef {hmap}[HM-rep]{Heat map}
\acrodef {rgrid}[RG-rep]{Refined grid}
\acrodef {CNN}[CNNs]{Convolutional Neural Network}
\acrodef {GAN}[GAN]{Generative Adversarial Net}
\acrodef {DANN}[DANN]{Domain Adversarial Neural Networks}
\acrodef {GRL}[GRL]{Gradient Reversal Layer}
\acrodef {dint}[int-dist]{intermediate discrimination}
\acrodef {dout}[out-dist]{output discrimination}
\acrodef {ET}[ET]{explicit transformation}
\acrodef {FP}[FP]{false positives}
\acrodef {FN}[FN]{false negatives}
\acrodef {TP}[TP]{true positives}

\acrodef {SNR}[SNR]{signal-to-noise ratio}
\acrodef {FC}[FC]{fully-connected}

\usepackage{colortbl}
\usepackage{hhline}
\usepackage{enumitem}
\usepackage{booktabs}
\newcommand{\ts}{\textsubscript}
\newcolumntype{C}[1]{>{\centering\let\newline\\\arraybackslash\hspace{0pt}}m{#1}}

\let\svthefootnote\thefootnote
\newcommand\freefootnote[1]{%
  \let\thefootnote\relax%
  \footnotetext{#1}%
  \let\thefootnote\svthefootnote%
}

\title{Data-Efficient Framework for Real-World \\ Multiple Sound Source 2D Localization}
\name{
\begin{tabular}{c}
Guillaume Le Moing$^{1,2,\dagger}$\thanks{$\dagger$ Work performed during internship at IBM Research Tokyo},
Phongtharin Vinayavekhin$^{1}$, Don Joven Agravante$^{1}$, Tadanobu Inoue$^{1}$ \\
Jayakorn Vongkulbhisal$^{1}$, Asim Munawar$^{1}$, Ryuki Tachibana$^{1}$
\end{tabular}}

\address{
  $^{1}$IBM Research, Tokyo, Japan\\
  $^{2}$Inria, École normale supérieure, CNRS, PSL Research University, Paris, France\\
}

\begin{document}
\ninept

\maketitle

\begin{abstract}
Deep neural networks have recently led to promising results for the task of multiple sound source localization.
% Yet, they require a lot of data for training, which needs to cover a variety of acoustic conditions and microphone configurations.
Yet, they require a lot of training data to cover a variety of acoustic conditions and microphone array layouts.
One can leverage acoustic simulators to inexpensively generate labeled training data.
However, models trained on synthetic data tend to perform poorly with real-world recordings due to the domain mismatch.
% Moreover, learning different microphone configurations highly complexifies the localization task due to the curse of dimensionality.
Moreover, learning for different microphone array layouts makes the task more complicated due to the infinite number of possible layouts.
We propose to use adversarial learning methods to close the gap between synthetic and real domains.
Our novel ensemble-discrimination method significantly improves the localization performance without requiring any label from the real data.
% Furthermore, we embed an explicit transformation layer in our architecture which enables the model to be trained with data from specific microphone array layouts while generalizing well to various unseen layouts during inference.
Furthermore, we propose a novel explicit transformation layer to be embedded in the localization architecture.
It enables the model to be trained with data from specific microphone array layouts while generalizing well to unseen layouts during inference.
\end{abstract}

\begin{keywords}
2D sound localization, multiple sound sources, deep learning, adversarial domain adaptation, novel network layer
\end{keywords}

\freefootnote{
\copyright~2021 IEEE. Personal use of this material is permitted. Permission from IEEE must be obtained for all other uses, in any current or future media, including reprinting/republishing this material for advertising or promotional purposes, creating new collective works, for resale or redistribution to servers or lists, or reuse of any copyrighted component of this work in other works.
}

\section{Introduction}
\label{sec:intro}

\ac{SSL} aims at estimating the pose/location of sound sources.
With the increasing popularity in installing smart speakers in home environments, source location provides additional knowledge that could enable a variety of applications such as monitoring daily human activities~\cite{Vacher2011}, speech enhancement~\cite{jeyasingh2018} and human-robot interaction~\cite{jour:csl:argentieri2015}.
% other than monitoring human speech
\ac{SSL} is an active research topic for which various signal-processing methods have been proposed~\cite{jour:csl:argentieri2015, jour:wcmc:cobos2017}.
These data-independent methods work well under strict assumptions~\cite{jour:wcmc:cobos2017}, e.g. high \ac{SNR}, known number of sources, and low reverberation.
Such ideal conditions hardly hold true in real-world applications and usually require special treatments~\cite{jour:signal:ma2006, jour:tasl:Alexandridis2018}.
%  conf:interspeech:guo2016}.
%conf:interspeech:netsch2014, 
Data-driven methods, and in particular deep learning, have recently outperformed classical signal-processing methods for various audio tasks
\cite{jour:ispl:salamon2017,tech:dcase:Inoue2018}
including \ac{SSL}~\cite{conf:icassp:pertila2017,
% conf:eusipco:adavanne2018,
conf:interspeech:he2018
% conf:icra:he2018,
% conf:icassp:takeda2016
% jour:sensors:diaz2018,
% conf:ica:pujol2019,
% jour:aslp:ma2017}.
}.

% Multiple network architectures have been proposed to localize sound sources.
% Network architectures have been proposed to localize sound sources for a variety of applications~\cite{conf:interspeech:he2018, conf:icassp:takeda2016, jour:wcmc:cobos2017}.
% The major advantage of these data-driven methods is their ability to adapt to challenging acoustic conditions and microphone array layouts.
% An advantage of these methods, apart from their ability to adapt to challenging 
% acoustic conditions and microphone configurations, is that they can be trained 
% to solve multiple tasks at the same time like simultaneous localization and 
% classification of sounds~\cite{conf:interspeech:he2018}.
% Yet, a significant downside is that they require lots of training data, which is expensive to gather and label~\cite{conf:icassp:He2019}.
% In addition, a pain point in applications which require multiple microphone arrays is that the localization network is only suited for array layouts it was trained for~\cite{conf:mmsp:lemoing2019, jour:sensors:diaz2018, conf:mlsp:vesperini2016}.

An advantage of these data-driven methods is their ability to adapt to challenging acoustic conditions.
% Yet, a downside is that they require lots of training data, which is expensive to gather and label~\cite{conf:icassp:He2019}.
Yet, they require a large amount of training data, which is expensive to gather and label~\cite{conf:icassp:He2019}.
% Acoustic simulators are an appealing solution as they can generate high-quality labeled samples in quantity.
Acoustic simulators are an appealing solution as they can abundantly generate high-quality labeled datasets.
% Although they can cover various layout configurations, this approach still suffers from the curse of dimensionality.
% Moreover, models trained on synthetic data (source domain) may undergo a critical drop of performance when exposed to real-world data (target 
However, models trained on synthetic data as a source domain can suffer from a critical drop in performance when exposed to real-world data as the target domain.
This is due to acoustic conditions that are outside the distribution of the synthetic dataset~\cite{jour:jrm:yalta2017,conf:icassp:carlo2019}, resulting in a \textit{domain shift}~\cite{jour:springer:ben-david2010}.

Recently, there have been several attempts to tackle domain adaptation for \ac{SSL}.
Unsupervised methods using entropy minimization of the localization output~\cite{conf:icassp:Takeda2017, conf:icassp:Takeda2018} have been proposed.
However, such methods are not suitable to our problem because entropy 
minimization encourages the prediction of a single 
source, whereas we must cater to multiple sources.
In this context, He et al.~\cite{conf:icassp:He2019} proposed two adaptation 
methods compatible with multiple sources.
First, a weakly supervised method conditioned on the expected number of sources, but this number is not known in our case.
Second, an unsupervised method based on \ac{DANN}~\cite{jour:jmlr:Ganin2015} which intends to align latent feature distributions for synthetic and real-world domains, but it was reported that the method did not improve the localization performance.
Still, adversarial methods, like \ac{DANN}~\cite{jour:jmlr:Ganin2015}, are popular outside \ac{SSL}, e.g. in computer vision~\cite{jour:coor:Vu2018}, and for other audio tasks such as acoustic scene
classification~\cite{conf:icassp:wei2020} and sound event detection~\cite{workshop:dcase:Gharib2018}.
Since it is not clear whether adversarial methods are suitable for \ac{SSL}, we present the results of extensive experiments with such methods.

Another advantage of these data-driven methods is that a single model can be trained to work with various microphone array layouts.
However, the model only performs well with the layouts that existed in the training data~\cite{conf:mmsp:lemoing2019, jour:sensors:diaz2018, conf:mlsp:vesperini2016}.
For different layouts, an additional training dataset is required.
Hence, the model cannot be used in applications with moving microphone arrays or for which the spatial layout of arrays is not known beforehand.
To the best of our knowledge, no prior work has tackled this issue so far.

\begin{figure*}[ht]
    \centering
    \includegraphics[width=0.92\textwidth]{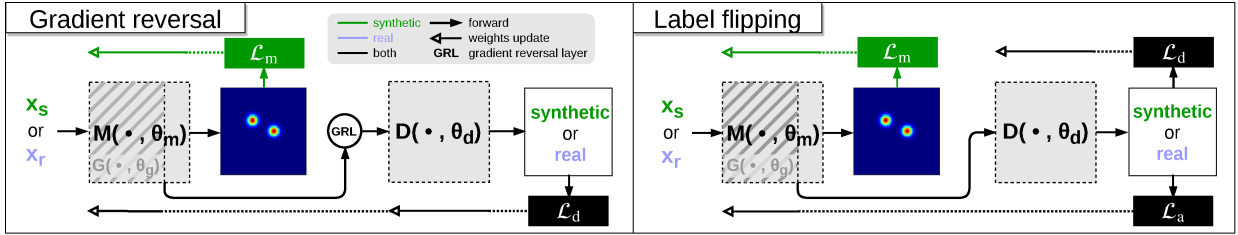}
    % \includegraphics[width=\linewidth, natwidth=880, natheight=167]{figures/label_flipping_gradient_reversal_v3.pdf}
    % \caption{Outline of the domain adaptation framework for the two studied methods in the case of intermediate discrimination.}
    \caption{Outline two domain adaptation methods with intermediate discrimination.}
    \label{fig:label_flipping_gradient_reversal}
    \vspace{-1.0em}
\end{figure*}

In this work, we address two critical issues of data-driven methods for \ac{SSL} : the domain shift and the difficulty to generalize to unseen layout configurations. Our contributions are two-fold:
% say the name of both method; and say we state the difference
% say we proposed a novel method based on ensemble method and describe some details
\begin{itemize}[wide,labelwidth=!,labelindent=0pt, noitemsep, topsep=0pt]
%\item We compare two domain adaptation methods based on adversarial learning, gradient reversal and label flipping. In both cases, we propose a novel ensemble discrimination extension, with discriminators at different levels of the localization model (intermediate and output space). Through extensive experiments, we show that our ensemble discrimination method outperforms the standard adversarial learning approaches. Although it does not require any label from the real-world data, our method can perform nearly as well as localization models that were trained in a supervised fashion on labeled real-world data.
\item We investigate two domain adaptation methods based on adversarial learning: gradient reversal and label flipping.
For both methods, we propose an ensemble-discrimination method that combines discriminators at different levels of the localization model: the intermediate and output spaces.
Through extensive experiments, we show that our ensemble-discrimination method outperforms the standard adversarial learning approaches and, without using any label for the real-world data, can perform nearly as well as localization models that were trained in a supervised fashion on labeled real-world data.
% Although it does not require any label from the real-world data, our method can perform nearly as well as localization models that were trained in a supervised fashion on labeled real-world data.

% say how much effort we need on processing a result/ GPU hours (didn't stored the timing; GPU is a bit risky to show)
% (c) We evaluate our proposed method extensively on data captured from a real environment.
% For each method, we ensure the consistency of the results by performing the experiments on different random seeds and reporting their mean and variance.
\item We introduce a novel neural network layer, the explicit transformation layer, which makes the localization model invariant to the microphone array layout.
Specifically, it allows the network to explicitly use the relative poses between microphone arrays which, in turn, makes the network generalize better to data in unseen layouts during inference.
Experiments show that the proposed layer learns more efficiently from varied layouts than baseline methods which process the layout implicitly.
Moreover, only our method is able to localize sources from new layouts when trained with a fixed layout.

% our proposed method allows the model to be trained with the data from specific microphone array layouts while generalizing well to data in various unseen layouts during inference.
% This also leads to a significant decrease in the amount of training data required for \ac{SSL} task where the layouts are varied during inference.
% We outperform by a large margin baseline methods for which the layout configuration is an additional input that the network needs to learn.
% Moreover, we show that our method can generalize from a fixed microphone array layout to unseen layouts while the baselines cannot.
\end{itemize}

\section{Domain Adaptation for Multiple Sound Source 2D Localization}
\label{sec:adaptation-method}

% Testing on real data a deep learning model that was trained on synthetic data, can result in a performance drop due to the
Training a deep learning model on synthetic data and testing it on real data usually results in a performance drop due to the
% distributions being different, that is, the simulator is not perfect.
distributions being different, that is, the simulator cannot match all real configurations.
However, improving the simulation model is often too difficult. A more 
practical approach is randomizing the 
simulator parameters~\cite{jour:iros:tobin2017}. This \textit{expands} the
source domain in the hopes of covering the 
target domain, e.g.\ parameters such as noise 
and reverberation can be randomized to achieve this effect.
% Assuming we can no longer improve the simulator and the model performance on real data is still insufficient, we need to use domain adaptation techniques~\cite{jour:springer:ben-david2010}.
Assuming we can no longer improve the simulator and the model performance on real data is still insufficient, domain adaptation methods~\cite{jour:springer:ben-david2010} can be used.
This paper focuses on enhancing these methods, in particular those that use adversarial learning, using an ensemble of discriminators.

We start by describing the base task of learning to predict the 
location of multiple simultaneous sound sources in a 2D horizontal plane.
To do that, we extract spectral features $\mathcal{X}$ from the sound recorded by a 
set of microphones. Then, we train a localization model $M$, with parameters $\theta_m$, 
to map spectral features to localization heatmaps $\mathcal{Y}$: discretized grids with Gaussian distributions centered at the source locations. 
Training the model amounts to solving:
\begin{equation}
	\min_{\theta_m}\sum_{(x,y)\in\mathcal{X}\times\mathcal{Y}}\mathcal{L}_\text{m}(M(x), y)
	.
    \label{eq:loc-obj}
\end{equation}
% The localization loss $\mathcal{L}_\text{m}$ is the mean squared error. Readers can find more details in \cite{conf:mmsp:lemoing2019} where this framework was introduced.
where the localization loss $\mathcal{L}_\text{m}$ is the mean squared error. More details on the framework can be found in a previous study~\cite{conf:mmsp:lemoing2019}.

We wish to augment (\ref{eq:loc-obj}) to perform well on real data 
$\mathcal{X}_r$ while only 
having labels $\mathcal{Y}_s$ for the synthetic data $\mathcal{X}_s$.
Without labels $\mathcal{Y}_r$, (\ref{eq:loc-obj}) cannot be used directly to learn
the mapping from real-world sound features to location cues.
This is an \textit{unsupervised domain adaptation} problem. To solve 
this, we use adversarial learning to make 
features generalizable and indistinguishable between the \textit{synthetic} and 
\textit{real} domains.
A discriminative model $D$, with parameters $\theta_d$, is plugged at a 
given layer of the localization neural network model (see \cref{fig:label_flipping_gradient_reversal}).
We denote the submodel of the localization model up to this layer as $G$, 
with parameters $\theta_g$ (a subset of $\theta_m$).
Here, $M$ is always trained with (\ref{eq:loc-obj}), using only 
synthetic labeled data, and 
$D$ is trained to assign domain class labels ($1$ for synthetic and $0$ 
for real) by using a \ac{BCE} loss ($\mathcal{L}_\text{d}$).
Meanwhile, $G$ attempts to generate features that cannot be distinguished by $D$.
This can be formalized as a minimax objective~\cite{conf:nips:goodfellow2014}:
\begin{equation}
  \small
  \min_{\theta_d}\max_{\theta_g}
  \sum_{x_s\in\mathcal{X}_s}
  \mathcal{L}_\text{d}(D(G({x_s})),1)
  +
  \sum_{x_r\in\mathcal{X}_r}
  \mathcal{L}_\text{d}(D(G({x_r})), 0)
  .
  \label{eq:minimax-obj}
\end{equation}
%\begin{multline}
%	\min_{\theta_d}\max_{\theta_g}\sum_{x_s\in\mathcal{X}_s}\mathcal{L}_\text{DIS}(D(G({x_s})), 1)
%	\\	
%	+
%	\sum_{x_r\in\mathcal{X}_r}\mathcal{L}_\text{DIS}(D(G({x_r})), 0)
%	.
%    \label{eq:minimax-obj}
%\end{multline}
% \subsubsection{Adversarial Learning in Practice}
To implement (\ref{eq:minimax-obj}), it has to be formalized as a cost-function minimization.
There are two methods for solving this in practice.

\textbf{Gradient reversal} The \ac{GRL}~\cite{jour:jmlr:Ganin2015} is a weight-free layer, placed between $G$ and $D$. It behaves as the 
identity during forward pass and negates the gradients during 
backward pass; changing the $max$ to a $min$ for optimizing weights $\theta_g$.

\textbf{Label flipping} is another method that is commonly used in the 
adversarial learning community. It involves decomposing the minimax 
objective (\ref{eq:minimax-obj}) into two minimization problems:
\begin{equation}
	\min_{\theta}
	\sum_{x_s\in\mathcal{X}_s}
	\mathcal{L}_\text{d}(D(G({x_s})), \alpha)
	+
	\sum_{x_r\in\mathcal{X}_r}
	\mathcal{L}_\text{d}(D(G({x_r})), \beta)
	,
    \label{eq:dis-obj}
\end{equation}
where the equation changes depending on the setting for $(\theta,\alpha,\beta)$.
To optimize $\theta_d$, we set it to $(\theta_d,1,0)$ and refer to it as
(\ref{eq:dis-obj}a).
To optimize $\theta_g$, we flip the 
labels by using the setting $(\theta_g,0,1)$ and refer to it 
as (\ref{eq:dis-obj}b).
% Furthermore, to help illustrate this difference in 
% \cref{fig:label_flipping_gradient_reversal}, we also use $\mathcal{L}_\text{a}$ 
% (adversarial loss) instead of $\mathcal{L}_\text{d}$ when using 
% (\ref{eq:dis-obj}b).
To help illustrate this difference in \cref{fig:label_flipping_gradient_reversal}, we replace label $\mathcal{L}_\text{d}$ with $\mathcal{L}_\text{a}$, an adversarial loss, when using (\ref{eq:dis-obj}b).

% \textbf{Comparison between both methods} 
Although both gradient reversal and label flipping are methods that intend to solve
(\ref{eq:minimax-obj}), there are important differences.
First, gradient reversal requires one forward-backward pass in $D$ at each update step whereas label 
flipping requires two passes, that is, one 
for each objective (\ref{eq:dis-obj}a), (\ref{eq:dis-obj}b).
% Second, they are not equivalent in terms of gradient computation.
Second, their gradient computation differs in the magnitude at each update step.
For label flipping, the weights update for $G$ and $D$ follows the same dynamic 
with respect to their objective so that the update is larger the farther
it is from the optimum.
In contrast, the dynamic is inverted for gradient reversal for updates 
on $\theta_g$. This results in smaller updates farther 
from the optimum, slowing down convergence.
This can cause $D$ to converge faster than $G$, which may result in $D$ not being able to 
provide sufficient gradients to improve $G$~\cite{conf:nips:goodfellow2014}.
Although we present this basic analysis, stable adversarial learning is still an active research topic.
Therefore, we compare both methods with extensive empirical results.

\textbf{Ensemble of Discriminators}
% We discuss two potential discrimination levels where the methods discussed above can be applied.
To describe our proposed ensemble discrimination method, we must first present what discrimination levels in adversarial learning mean.
A discrimination level refers to the layer of $M$ at which we conduct adversarial learning.
Although this can be continuously moved, we opt to take only the two extreme levels of discrimination.
The first is \ac{dint}, where we place the discriminator, $D$, right after the \textit{encoder} of the \ac{SSL} model in~\cite{conf:mmsp:lemoing2019}.
In this case, the submodel $G$ is the encoder.
We do not go further into the encoder to allow enough \textit{capacity} for $G$ to learn the domain independent features.
The second is \ac{dout}, where we place $D$ after the output layer such that $G$ is all of $M$.
Note that $\mathcal{Y}_s$ and $\mathcal{Y}_r$ distributions being similar (number and 2D location of sources) is a prerequisite for the success of this strategy.
If not, $D$ can learn the dissimilarities and the generator will unwantedly distort its predictions to satisfy (\ref{eq:minimax-obj}).
Constraining the output is common for domain adaptation in \ac{SSL} with, for example, entropy methods~\cite{conf:icassp:Takeda2017, conf:icassp:Takeda2018} or weak supervision~\cite{conf:icassp:He2019}.
However, such methods can degrade the performance by boosting incorrect low confidence predictions~\cite{conf:icassp:He2019} resulting in more false positives.
Using \ac{dout} has this same concern.
% Lastly, we propose a type of novel \textit{ensemble} method by using both \ac{dint} and \ac{dout}.
Lastly, our \textit{ensemble-discrimination} method uses both \ac{dint} and \ac{dout}.
For this, the two discriminators are trained independently, together with one common localization model $M$.
We expect that each discriminator will enhance $M$ in different aspects and lead to improved results.
% Each discriminator would improve $M$ in different aspects, which we expect to lead to better overall performance.
% Our intuition is that doing so would improve M in different ways leading to better results.

\section{Explicit Transformation Layer}
\label{sec:transformation-method}

\begin{figure}
    \centering
    \includegraphics[width=\columnwidth]{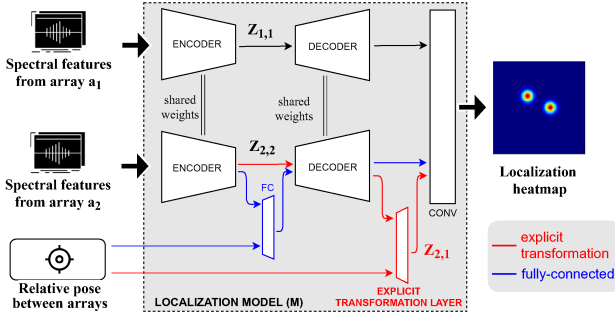}
	\caption{Layout generalization methods: \acs{ET} layer and two baselines.}
    \label{fig:et}
\end{figure}

% We describe our \ac{ET} layer for layout invariant localization (see~\cref{fig:et}).
% We use a layout with two microphone arrays ($a_1$ and $a_2$), but the extension to layouts with more arrays is straightforward.
We introduce a novel \ac{ET} layer for layout invariant \ac{SSL} (see~\cref{fig:et}).
Here, we describe the use of the \ac{ET} layer in a layout with two microphone arrays, $a_1$ and $a_2$,but the extension to more arrays is straightforward.
Let tensor $Z_{i, j}$ of size $W$x$H$x$C$ be the output of decoder of $a_i$ in the coordinate system of $a_j$, where $W$, $H$ and $C$ are the width, height and number of feature maps.
$Z_{1, 1}$ and $Z_{2, 2}$ are the output of the decoder after passing raw sound of each array through the network of~\cite{conf:mmsp:lemoing2019}.
The role of \ac{ET} is to transform $Z_{2, 2}$ to $Z_{2, 1}$.
Let $Z_{i,j}^k$ of size $W$x$H$x$1$, be the $k$\textsuperscript{th} feature map of tensor $Z_{i,j}$.
Let $(t_x, t_y, \theta)$ be the relative pose of $a_2$ from $a_1$ in a 2D plane, the 2D Euclidean transformation from $a_2$ to $a_1$ is defined by:
\begin{equation}
   {}_{1}H_{2} = \
   \begin{bmatrix}
   \cos{\theta} & -\sin{\theta} & t_{x}\\
   \sin{\theta} & \cos{\theta} & t_{y} \\
   0 & 0 & 1
   \end{bmatrix}.
    \label{eq:etl}
\end{equation}
Such that the relation of locations can be computed as $(x_1, y_1, 1)^{T} = {}_{1}H_{2}(x_2, y_2, 1)^{T}$, or equivalently $(x_2, y_2, 1)^{T} = {}_{1}H_{2}^{-1}(x_1, y_1, 1)^{T}$, where $(x_1, y_1)$ is a position in $Z_{2,1}^k$ and $(x_2, y_2)$ is a position in $Z_{2,2}^k$.

Now that we have this correspondence, we can compute the values of $Z_{2,1}^k$.
First, for each position in $Z_{2,1}^k$, we find the corresponding point in space $(x_2', y_2')$ based on the transformation above.
This has a floating-point value.
We then take the values of the $n$ closest discrete positions in $Z_{2,2}^k$ and compute the value of $Z_{2,1}^k$ using the bilinear interpolation method.
% For edge cases where there is no corresponding point, zero is used instead.
This is done for all positions of $Z_{2,1}^k$.
When applied to all feature maps of $Z_{2, 2}$, we obtain $Z_{2,1}$ which is the desired output of the \ac{ET} layer.

% intuition on what it does
\ac{ET} is a specialized form of spatial transformer network~\cite{conf:nip:jaderberg2015}.
It is a weight-free layer which is computed according to the known relative position between the two microphone arrays.
Intuitively, the role of the \ac{ET} layer is to transform the decoder output of a target microphone array from its own coordinate system to the coordinate system of the reference microphone array.
In addition, the gradient can back-propagate through the \ac{ET} layer during training.

\section{Experiments}

\subsection{Dataset}

% Our data-collection setup was completely described 
% in~\cite{conf:mmsp:lemoing2019} and we recall the main points here.
Our data-collection setup was extensively described in the previous study~\cite{conf:mmsp:lemoing2019}, but the main points are recalled here.
Experiments are conducted in a $6 \text{m} \times 6 \text{m}$ area with two 
microphone arrays.
Music clips from the classical and funk genres are used for recording 
training, validation and testing data splits. An additional testing dataset from the jazz genre is recorded to verify the generalizability of the model.
One or two sources can be active simultaneously, and sources are at least $2 \text{m}$ apart to ease the evaluation.

We synthesize data with Pyroomacoustics~\cite{conf:icassp:scheibler2018}, 
where we recreated the real room configuration.
We generate three training datasets of $10^6$ samples:
\textbf{TrainS}, an anechoic chamber with a low-noise setting and fixed array layout,
% \textbf{TrainS+}, same as TrainS with augmentation (random size for the room and signal-to-noise ratio), and
\textbf{TrainS+}, same as TrainS but room size and \ac{SNR} are randomized, and
\textbf{TrainSa}, same as TrainS with various microphone array layouts.
Another synthetic dataset of size $5\times10^3$ with various microphone array layouts, \textbf{TestSa}, is captured for testing.
We make sure there is no overlap between TrainSa and TestSa in terms of layout configurations.

We also record real-world data.
To ease labeling,
we use an augmentation method~\cite{conf:mmsp:lemoing2019},
wherein we capture one active source at a time and use
sound superposition to generate samples with multiple sources. $0.16$s audio samples are then extracted.
%In each $1 \text{m} \times 1 \text{m}$ cell of the area, we randomly selected and recorded $7, 1, 1$
%positions for training, validation and 
%testing 
%data, respectively.
\textbf{TrainR}, \textbf{ValidR}, \textbf{TestRc} and \textbf{TestRj} are the datasets for training, validating and testing respectively.
All of them are made of classical-funk music clips, 
except for TestRj which is made of jazz.
Training sets have $10^6$ samples, while other sets have $5\times10^3$ samples.

\subsection{Experimental Protocols}

% \textbf{Training} 
Our localization model, $M$, has the \textit{encoder-decoder} architecture \cite{conf:mmsp:lemoing2019} and takes as input \ac{STFT} features.
We conduct independent experiments for adversarial adaptation methods and generalization to arbitrary layouts methods.

For adversarial adaptation, input features are processed in individual encoders for each array, merged and then decoded together.
Discrimination is conducted on the merged encoded features for \ac{dint}, and 
localization heatmaps for \ac{dout}.
Discriminator \ac{dint} consists of $4$ dense layers while \ac{dout} is composed of $4$ 
convolutional layers followed by $3$ dense layers.
A ReLu activation is used after each layer, except for the last one which is a sigmoid.

For layout generalization, we use individual decoders for each array and we use channel concatenation followed by $3$ convolutional layers to merge the two streams after the decoding part (see \cref{fig:et}).
% The real and imaginary parts of the \ac{STFT} of 16kHz audio samples are
% extracted with Hamming window of 32 ms with $50\%$ overlap, discarding the 
% zero-frequency component, are used as the input features.

%Each layer is coupled with a ReLu activation function except the last one uses 
% a Sigmoid activation.
%

% \textbf{Evaluation Metrics} 
Our evaluation metrics are precision, recall and F1-score.
Source coordinates are extracted from the output heatmap as predicted 
keypoints~\cite{conf:mmsp:lemoing2019}, and then matched with the groundtruth keypoints with a resolution threshold 
of 0.3m on synthetic data and 1m on real data.
Based on the matches, we count the \ac{TP}, \ac{FP} and \ac{FN}, from which we compute the F1-score.
The \ac{RMSE} of the matches is provided as an additional indicator of the quality.

% \textbf{Methods for Comparison} 
Methods are trained for $200$ epochs with ADAM optimizer, initial learning rate of $10^{-4}, \beta=(0.5, 0.999)$ and batch size of $256$.
Adversarial methods being notoriously unstable, we report the mean and standard deviation of 5 independent trainings with different random seeds.
For layout generalization methods, we report one run.
\begin{itemize}[wide,labelwidth=!,labelindent=0pt, noitemsep, topsep=0pt]

\item \textbf{\textit{Lower bound}} method \textbf{S} is a supervised 
method trained on labeled synthetic data (TrainS).
It represents the minimum performance without any knowledge of the target 
domain.

\item \textbf{\textit{Upper bound}} method \textbf{R} is a supervised method 
trained on labeled real-world data (TrainR).
It is an approximation of the maximum performance~\cite{jour:springer:ben-david2010}, when a lot of efforts is dedicated to label real-world data which we would like to avoid in 
practice.

% \item \textbf{\textit{Others}} refers to methods modifying the training data.
% Method \textbf{R\&S} uses data augmentation by using TrainR and TrainS, both 
% with labels, to train $M$.
\item \textbf{\textit{Others}} refers to methods that use modified training data.
Method \textbf{R\&S} uses TrainR and TrainS, both with labels, to train $M$.
It sacrifices data quality by mixing domains, but has the same data quantity used in the adversarial methods.
The main burden is that it requires labels for the real-world data.
Method \textbf{S+} uses domain randomization~\cite{jour:iros:tobin2017} by using
TrainS+ instead of TrainS in method S.

\item \textbf{\textit{Adversarial Adaptation}} methods are the variations of 
\textit{unsupervised domain adaptation methods using adversarial learning}.
% that we described.
These are trained on TrainS with labels and TrainR without labels.
These methods are: 
\textbf{G\ts{r}int}, \textbf{L\ts{f}int}, 
\textbf{G\ts{r}out}, \textbf{L\ts{f}out},
and ensemble methods 
\textbf{G\ts{r}intG\ts{r}out} and 
\textbf{L\ts{f}intL\ts{f}out}.
\textit{G\ts{r}} and \textit{L\ts{f}},
refer to gradient reversal and label flipping respectively, and
\textit{int} and \textit{out} are short for the discrimination levels
\ac{dint} and \ac{dout}.
We also considered \textbf{G\ts{r}intG\ts{r}out+} and 
\textbf{L\ts{f}intL\ts{f}out+} which uses TrainS+ instead of TrainS.

% \item \textbf{\textit{Layout Generalization}} methods comprise of a the proposed \textbf{Explicit Transformation} layer and two baselines.
% First, \textbf{Plain Encoder-Decoder} the plain network without \ac{ET} layer.
% Second, \textbf{Fully-connected} where the relative pose between arrays is processed through a fully-connected layer and resulting features are concatenated with the output of the array encoder (see \cref{fig:et}).

\item \textbf{\textit{Layout Generalization}} methods are composed of two baselines and the proposed method, shown in~\cref{fig:et}.
	(a) \textbf{Plain Encoder-Decoder} (black) is the plain network without any additional input or layer.
	(b) \textbf{Fully-connected} (blue) receives the relative pose between arrays as input, processes it through a \ac{FC} layer, and concatenates the resulting features to the output of the encoder.
	(c) \textbf{Explicit Transformation} (red) uses the relative pose to define an \ac{ET} layer which then transforms the output of the array decoder.

\end{itemize}

\subsection{Results and Discussion}

\begin{figure}%
    \centering
    \includegraphics[width=7.5cm]{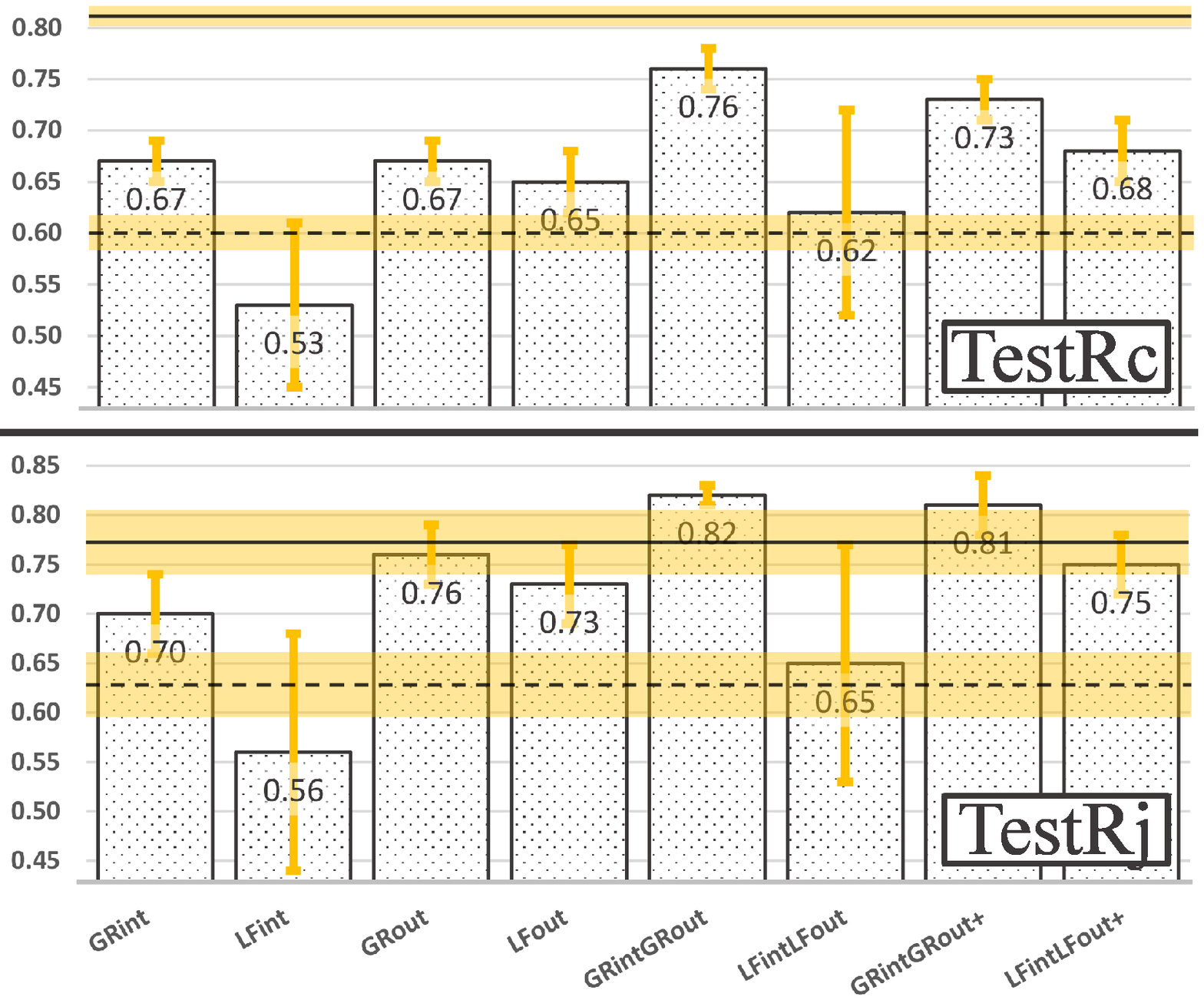}
    \qquad
    \includegraphics[width=7.0cm]{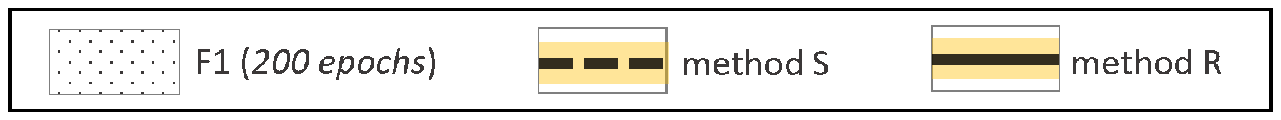}%
    \caption{F1-Scores of adversarial adaptation methods compared with upper-bound and lower-bound methods.}%
    \label{fig:f1score}%
    \vspace{-1.0em}
\end{figure}

\begin{table}
\footnotesize\addtolength{\tabcolsep}{-4pt}
\begin{center}
\caption{Domain adaptation results on additional metrics.}
\label{tab:results0a_200} %https://tex.stackexchange.com/a/59805
\begin{tabular}{C{1cm}C{2cm}C{0.75cm}C{0.75cm}C{0.75cm}C{0.75cm}C{0.75cm}C{0.75cm}}\toprule
\multicolumn{2}{C{3cm}}{(\textit{Test Data})}                                         & \multicolumn{6}{C{4.5cm}}{\textbf{TestRc (classical \& funk)}}                                                                                       \\ 
\cmidrule{3-8}
\multicolumn{2}{C{3cm}}{\textbf{\textit{Method}}}                                                       & \multicolumn{2}{C{1.5cm}}{\textbf{RMSE $\downarrow$}} & \multicolumn{2}{C{1.5cm}}{\textbf{Precision $\uparrow$}} & \multicolumn{2}{C{1.5cm}}{\textbf{Recall $\uparrow$}}  \\ 
\cmidrule{1-8}
Low B.                                                               & S                    & \multicolumn{2}{C{1.5cm}}{.51$\pm$.01}                 & \multicolumn{2}{C{1.5cm}}{.80$\pm$.03}                    & \multicolumn{2}{C{1.5cm}}{.48$\pm$.03}                  \\ 
\cmidrule{1-8}
Up B.                                                                & R                    & \multicolumn{2}{C{1.5cm}}{.34$\pm$.01}                 & \multicolumn{2}{C{1.5cm}}{.95$\pm$.03}                    & \multicolumn{2}{C{1.5cm}}{.71$\pm$.03}                  \\ 
\cmidrule{1-8}
\multirow{2}{*}{Others}                                              & R\&S                   & \multicolumn{2}{C{1.5cm}}{.33$\pm$.02}                 & \multicolumn{2}{C{1.5cm}}{.96$\pm$.03}                    & \multicolumn{2}{C{1.5cm}}{.71$\pm$.04}                  \\ 

                                                                     & S+                   & \multicolumn{2}{C{1.5cm}}{.44$\pm$.00}                 & \multicolumn{2}{C{1.5cm}}{.84$\pm$.01}                    & \multicolumn{2}{C{1.5cm}}{.57$\pm$.01}                  \\ 

\cmidrule{1-8}
\multirow{8}{*}{\begin{tabular}[c]{@{}c@{}}Adv.\\Adap.\end{tabular}} & GRint~\cite{jour:jmlr:Ganin2015,conf:icassp:He2019}               & \multicolumn{2}{C{1.5cm}}{.48$\pm$.03}             & \multicolumn{2}{C{1.5cm}}{.81$\pm$.05}                  &\multicolumn{2}{C{1.5cm}}{ .57$\pm$.04}                    \\ 

                                                                     & LFint~\cite{conf:nips:goodfellow2014, conf:cvpr:tsai2018}        & \multicolumn{2}{C{1.5cm}}{ .50$\pm$.03}              &\multicolumn{2}{C{1.5cm}}{ .71$\pm$.06}             &\multicolumn{2}{C{1.5cm}}{ .44$\pm$.09}                    \\ 

                                                                     & GRout                &\multicolumn{2}{C{1.5cm}}{ .51$\pm$.02}              & \multicolumn{2}{C{1.5cm}}{.68$\pm$.02 }             &\multicolumn{2}{C{1.5cm}}{ .67$\pm$.02}          \\ 

                                                                     & LFout                & \multicolumn{2}{C{1.5cm}}{.51$\pm$.02}              &\multicolumn{2}{C{1.5cm}}{ .64$\pm$.04}                  &\multicolumn{2}{C{1.5cm}}{ .67$\pm$.02}                    \\ 

                                                                     & GRintGRout           & \multicolumn{2}{C{1.5cm}}{.47$\pm$.05}              &\multicolumn{2}{C{1.5cm}}{\textbf{.82$\pm$.05}}    &\multicolumn{2}{C{1.5cm}}{\textbf{.72$\pm$.03}}                     \\ 

                                                                     & LFintLFout           & \multicolumn{2}{C{1.5cm}}{.53$\pm$.03}              &\multicolumn{2}{C{1.5cm}}{ .65$\pm$.10}             &\multicolumn{2}{C{1.5cm}}{ .60$\pm$.09}          \\ 

                                                                     & GRintGRout+          &\multicolumn{2}{C{1.5cm}}{\textbf{.46$\pm$.02}}              &\multicolumn{2}{C{1.5cm}}{ .76$\pm$.07}              & \multicolumn{2}{C{1.5cm}}{.71$\pm$.05}         \\ 

                                                                     & LFintLFout+          &\multicolumn{2}{C{1.5cm}}{ .50$\pm$.02}              & \multicolumn{2}{C{1.5cm}}{.72$\pm$.09}              &\multicolumn{2}{C{1.5cm}}{ .65$\pm$.02}          \\ 
\bottomrule                 
\end{tabular}

%%%%%%%%%%%%%%%%%%
\begin{scriptsize}
\begin{center}
\vspace{-1.2em}
% Results for the 200\textsuperscript{th}  epoch of training;
\textit{Low B.}, \textit{Up B.} and \textit{Adv. Adap.} refer to lower bound, upper bound and adversarial adaptation respectively. $\uparrow$/$\downarrow$ denote that higher or lower is better.
\end{center}
\end{scriptsize}
%%%%%%%%%%%%%%%%%%

\vspace{-4.0em}

\end{center}
\normalsize
\end{table}

%% no adaptation
\textbf{Adversarial learning} Localization performance across all methods for TestRc is illustrated in~\cref{fig:f1score} and~\cref{tab:results0a_200}.
For our lower bound baseline \textbf{S}, there are not many 
detections (low recall) but they are mostly correct (high 
precision), and correct predictions have a low quality (high RMSE).
% r method
In comparison, method \textbf{R} which uses labeled real-world data, yielded
significantly better performance across all metrics. 
In theory~\cite{jour:springer:ben-david2010}, our domain adaptation methods should 
have results within this range.
% r s method
Method \textbf{R\&S} marginally improves performance 
compared to method \textbf{R} on TestRc.
% s+ method
Method \textbf{S+} shows a slight improvement over \textbf{S} but is still far 
from method \textbf{R} and \textbf{R\&S}.
% domain adaptation methods
Most domain adaptation methods, especially our ensemble-discrimination methods, improve in performance compared to the \textit{lower bound} method \textbf{S}.
In particular, the improvement in recall for some of our proposed methods compared to \textbf{S} is significant and leads to a higher F1-score as shown in~\cref{fig:f1score}.

%% inference on jazz
The results of domain adaptation seem to be even better on TestRj, even though the model is trained on classifical and funk excerpts.
We speculate that this is due to TrainS being more similar to TestRj in terms of spectral features.
This explains why lower bound \textbf{S} is higher for TestRj while upper bound \textbf{R} is lower.

% dint dout
% One of our proposed ensemble discrimination methods, \textbf{G\ts{r}intG\ts{r}out}, lead to the best results across all test sets.
Methods based on G\ts{r} tend to have better performance and lower variance compared to L\ts{f}.
In particular, one of our ensemble-discrimination methods, \textbf{G\ts{r}intG\ts{r}out}, leads to the best results across all test sets and performs almost as well as method \textbf{R} while not using labels for the real data.
From \cref{tab:results0a_200}, we can see that methods with \ac{dint} lead to a higher precision and a lower \ac{RMSE} while \ac{dout} mainly increases recall.
It shows that \ac{dint} improves the existing detections while \ac{dout} encourages more detections but at a higher risk of false positives. All in all, combining them gives the best results.
% adv and s+
Randomization (using TrainS+ instead of TrainS) did not seem to further improve the ensembling results.
While most unsupervised domain adaptation methods studied in this paper outperform \textit{lower bound} methods which is contrary to the results presented in~\cite{conf:icassp:He2019}, adversarial methods are notoriously difficult to train~\cite{conf:nips:lucic2014}.
The high variance of L\ts{f} in our results further supports this.
%% previous method
% Finally, we note that although most of our results are contrary to the results in~\cite{conf:icassp:He2019}, adversarial methods are notoriously difficult to train~\cite{conf:nips:lucic2014}. The high variance of L\ts{f} in our experimental results further supports this.

\textbf{Layout Generalization} \cref{tab:results0c} displays the results on layout generalization.
When trained with a fixed array layout (TrainS), \ac{ET} can generalize to new layouts fairly well while the FC baseline cannot.
Moreover, when traind with various layouts (TrainSa), the plain network which has to learn the pose implicitly from the sound signals performs poorly while \ac{ET} yields the best F1-score and RMSE.

\begin{table}
\footnotesize\addtolength{\tabcolsep}{-2pt}
\begin{center}
\caption{Performance of layout generalization.}
\label{tab:results0c} %https://tex.stackexchange.com/a/59805

\begin{tabular}{lllll}\toprule
\multicolumn{1}{C{3cm}}{(\textit{Test Data})}                & \multicolumn{4}{c}{\textbf{TestSa}}  \\
\cmidrule(lr){2-5}
\multicolumn{1}{C{3cm}}{(\textit{Train Data})}      & \multicolumn{2}{c}{\textbf{TrainS}} & \multicolumn{2}{c}{\textbf{TrainSa}} \\
\cmidrule(lr){2-3} \cmidrule(lr){4-5}
\multicolumn{1}{C{3cm}}{\textbf{\textit{Method}}}                   & F1-Score       & RMSE      & F1-Score       & RMSE       \\
\cmidrule{1-5}
Plain Encoder-Decoder                  & -            & -       & .35            & .17        \\
Fully-connected & .09            & .18       & .54            & .17        \\
Explicit Transformation                  & \textbf{.62}            & \textbf{.14}       & \textbf{.80}            & \textbf{.11} \\ \bottomrule
\end{tabular}

\end{center}
\normalsize

\vspace{-1.0em}

\end{table}

\section{Conclusion}
\label{sec:conclusion}

% In this paper we studied adversarial domain adaptation methods from synthetic to 
% real-world data for deep learning based multiple sound source 2D localization.
This paper proposed two solutions for data-efficient learning of multiple sound source 2D localization.
First, we studied adversarial adaptation methods. These methods, and in particular the proposed ensembling method, yield a significant improvement compared to solely training on synthetic 
data, at a much lower cost than training on labeled real-world data.
Second, we made the localization model invariant to the pose of the microphone arrays. The proposed method outperforms baseline approaches and greatly reduces the need for data since it can generalize from a few layouts to unseen layouts.

\bibliographystyle{IEEEbib}
\bibliography{reference}

\end{document}